\documentclass[%
 aip,
 amsmath,amssymb,
 reprint,%
]{revtex4-1}

\usepackage{graphicx}
\usepackage{dcolumn}
\usepackage{bm}

\usepackage[utf8]{inputenc}
\usepackage[T1]{fontenc}
\usepackage{mathptmx}
\usepackage{etoolbox}
\usepackage{subfigure}
\usepackage{caption, xcolor}

\makeatletter
\def\@email#1#2{%
 \endgroup
 \patchcmd{\titleblock@produce}
  {\frontmatter@RRAPformat}
  {\frontmatter@RRAPformat{\produce@RRAP{*#1\href{mailto:#2}{#2}}}\frontmatter@RRAPformat}
  {}{}
}%
\makeatother

\begin{document}

\preprint{AIP/123-QED}

\title{Development of portable cosmic-ray muon detector array for muography}
\author{Yunsong Ning}
\author{Yi Yuan}
\author{Tao Yu}
\affiliation{ 
School of Physics, Sun Yat-sen University, 510275 Guangzhou, China\\
Platform for Muon Science and Technology, Sun Yat-sen University, Guangzhou, China
}%
\author{Hongyu Chen}
\affiliation{%
University College London, London, United Kingdom
}%
\author{Chengyan Xie}
\affiliation{ 
School of Physics, Sun Yat-sen University, 510275 Guangzhou, China\\
Platform for Muon Science and Technology, Sun Yat-sen University, Guangzhou, China
}%
\author{Hui Jiang}
\affiliation{ 
State Key Laboratory of Particle Detection and Electronics, Experimental Physics Department, Institute of High Energy Physics, Chinese Academy of Sciences, Beijing 100049, China
}%
\author{Hesheng Liu}
\author{Guihao Lu}
\author{Mingchen Sun}
\author{Yu Chen}
\author{Jian Tang}
\email{tangjian5@mail.sysu.edu.cn}
\affiliation{ 
School of Physics, Sun Yat-sen University, 510275 Guangzhou, China\\
Platform for Muon Science and Technology, Sun Yat-sen University, Guangzhou, China
}%

\date{\today}

\begin{abstract}
As the multidisciplinary applications of cosmic-ray muons expand to large-scale and wide-area scenarios, the construction of cosmic-ray muon detector arrays has become a key solution to overcome the hardware limitations of individual detector. For muography, the array-based detector design enables fast-scanning of large target objects, allowing for rapid identification of density variation regions, which can improve the efficiency of tomography. This paper integrates scintillator detector technology with Internet of things~(IoT) technology, proposing a novel array networking model for nationwide deployment. The model enables long-distance data collection and distribution, laying the foundation for future multidisciplinary applications such as muography and other fields.
\end{abstract}

\maketitle

\section{Introduction}
Cosmic-ray muons are secondary particles produced mostly by the decay of ${\pi}$ and K, which result from collisions between cosmic rays and atmospheric nuclei~\cite{Bonomi:2020dmm}. As a naturally abundant source of muons, they exhibit high penetration capability and stability. Combined with the low cost and scalability of muon detectors, cosmic-ray muons have seen widespread applications in recent years in fields such as muography and geophysics~\cite{tanaka2023muography,bonechi2020atmospheric}.

\begin{figure}[htbp]
\centering
\includegraphics[width=0.8\hsize]{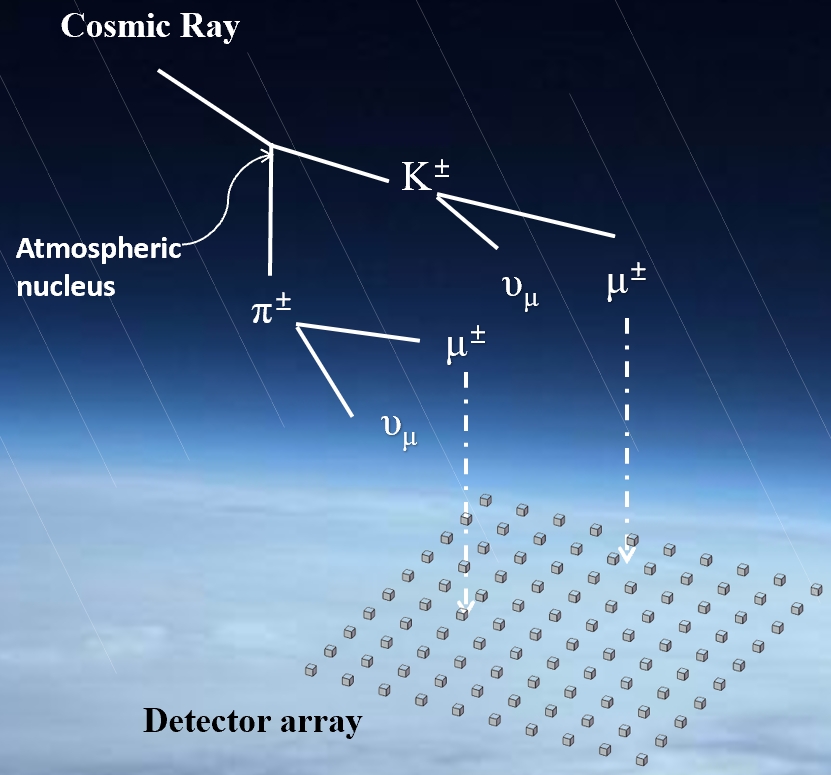}
\caption{\label{CRmu}The detector array can be used to detect cosmic ray atmospheric shower events}
\end{figure}

Muography is an emerging imaging technique developed in recent decades~\cite{kaiser2019muography}. It leverages the exceptional penetrating power of cosmic-ray muons to reconstruct 3D density profiles of objects by measuring muon transmission rates and angular scattering patterns. Pioneered in 1967 with the non-invasive scanning of Egyptian pyramids~\cite{alvarez1970search}, this technology has undergone significant advances in detector design. Modern implementations utilize multi-channel cosmic-ray muon detection systems to conduct spatially resolved scans, enabling applications across archaeology~\cite{morishima2023muography,d2020muography}, volcanic conduit mapping~\cite{tioukov2019first}, nuclear material non-deconstructive monitoring~\cite{RIGGI201816,6551201}, and tidal dynamics studies~\cite{tanaka2022periodic}. University of Tokyo has advanced volcanic monitoring through muography by performing tomographic scans of the Sakurajima and Asama volcanoes~\cite{leone2021muography,olah2018high}. By measuring density variations within volcanic structures before and after eruptions, they achieved a dynamic monitor of the volcanic activity. In 2020, the team developed the first muon-based monitoring system to track thickness and density changes of volcanic ash deposit, providing a high-precision assessment of eruption magnitudes~\cite{tanaka2020development}. Meanwhile, scientists from China have applied muography to mineral exploration, employing scintillator detectors to map cavities and fractures in Zaozigou gold mine~\cite{liu2024deep}, validating the feasibility of muography for the characterization of metallic ore-body. 

As multidisciplinary muon applications increasingly target large-scale and spatially extensive scenarios, the traditional single-detector system is confronted with inherent limitations due to prohibitive costs and technical barriers in scaling up the size of individual detector. To address this challenge, researchers have turned to modular detector array architectures, the methodology initially developed for cosmic-ray air shower detection, to drive transformative advancements in muography. 

The idea of a muon detector array was first proposed for monitoring atmospheric shower events~\cite{amenomori1990development}. The detection of cosmic-ray muons is instrumental in reconstructing the flux and energy distribution of cosmic rays and in exploring the mechanisms of neutrino production~\cite{Davier:2005id}. These contributions make muon a key component in multi-messenger astronomy~\cite{neronov2019introduction}, and the muon detector array is a common part of the air shower detection system. In the LHAASO experiment, the muon detection system comprises 1,171 underground water Cherenkov detectors integrated with PMTs, working in tandem with scintillator detectors to form the KM2A array~\cite{Li:2018vpy,LHAASO:2024kbg}. This hybrid configuration enables precise detection and discrimination of muon components within extensive air shower events. The muon detection results have significantly enhanced gamma-ray identification efficiency. Furthermore, they have led to the discovery of anomalies in the "knee" structure of the cosmic-ray energy spectrum at PeV energies and provided critical insights into the composition evolution of cosmic rays in the ultra-high-energy regime~\cite{LHAASO:2021gok,LHAASO:2023gne}.

The array-based detector design has significant potential in the field of muography. By deploying a distributed detector network, we will significantly reduce the imaging time and detector footprint while allowing sustained and high-resolution monitoring of geologically massive structures such as ore deposits and mountain systems. The scalability and adaptability of array-based systems thus overcome critical bottlenecks, unlocking the potential of the muography for long-term, large-area applications in resource exploration and geophysical studies.  

This article introduces a portable cosmic-ray muon detector based on plastic scintillators and silicon photomultiplier~(SiPM), which supports long-term monitoring of cosmic-ray muons in the field. The wireless module combined with the IoT technology embedded in the detectors enables them to form a detector network, paving the way for large-scale detection of cosmic ray muons and the development of subsequent detector arrays.

\section{Detector Structure and Function}
\subsection{Detector construction}

The portable cosmic-ray muon detector consists of two square plastic scintillators arranged in parallel upper and lower planes, paired with two Silicon Photomultipliers~(SiPMs). With the high penetration capability of cosmic muons, the system employs coincidence measurement techniques to identify muon signals. The plastic scintillators exhibit an emission peak at approximately 420 nm, with sizes of 10 cm ${\times}$ 10 cm ${\times}$ 1 cm ~\cite{zhong2025enhancing}, the recipe of which is optimized for specific operational requirements. The type of SiPMs used is MicroFJ-60035 from Onsemi. The SiPMs are optically coupled to the scintillators with liquid silicone grease to minimize light loss at the interface. 

\begin{figure}[htbp]
\centering
\includegraphics[width=1\hsize]{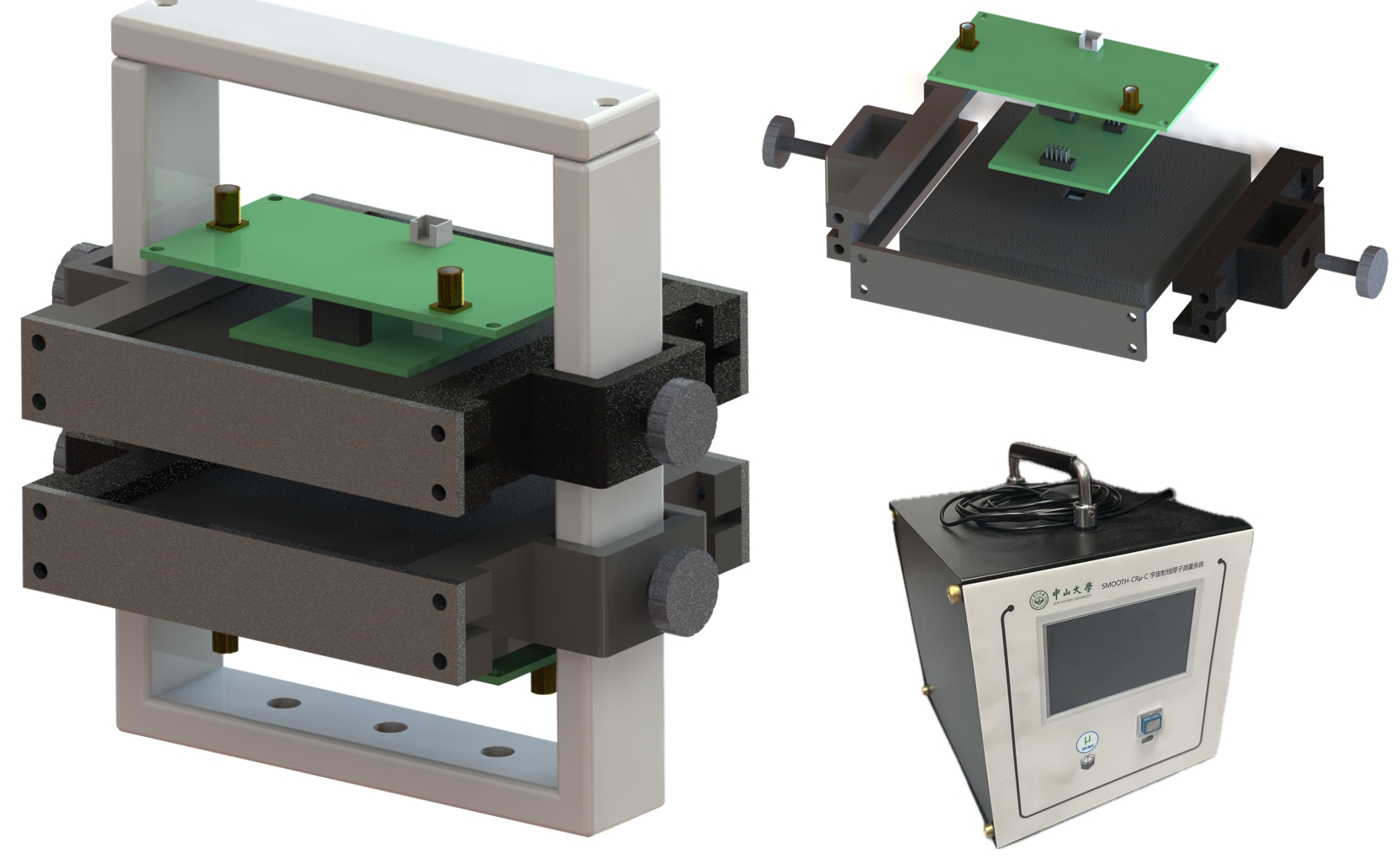}
\caption{\label{detector structure}The portable cosmic-ray muon detector and the inside structure. With the movable structure, the spatial acceptance angle of the detector can be adjusted.}
\end{figure}

To optimize light collection efficiency and minimize interference from ambient stray light, the scintillator is wrapped with a 125 ${\mu}$m aluminum foil reflector, followed by a black outer light-tight layer. In addition, the distance between two plastic scintillators can be easily changed with the support structure, providing an adjustable spatial acceptance.

The detector utilizes the ATmega2560 microcontroller on the Arduino platform to read and filter SiPM signals~\cite{axani2019physics}. To match the microcontroller's performance characteristics, an operational amplifier is integrated downstream of the SiPM to amplify and extend the signal, achieving a pulse width of approximately 400 ${\mu}$s. The circuit board operates a combination of a boost converter and a low-dropout regulator (LDO) to convert the external input power of 5–6 V into a stable supply of 29 V for the SiPMs. The low-power design enables the detector to operate on a lithium battery power, making it suitable for wild field applications.  

In a word, the portable cosmic-ray muon detector features a support structure made of aluminum alloy and stainless steel, with a total weight of less than 10 kg. Combined with its low-power lithium battery system, the detector is well suited for long-term field experiments.

\subsection{Detector functionality and performance}

The portable detector leverages the high extensibility of the Arduino platform, incorporating features such as GPS positioning and timing, local data storage, and real-time display of statistical results. These enhancements facilitate independent operation and deployment of the detector.

Designed for field experiments and array-based deployments, the detector system requires precise external timing and location data. A built-in GPS module provides timing synchronization and positioning through satellite navigation, delivering accurate reference coordinates for the detector. 

In addition, the detector utilizes a serial display screen to visualize cosmic-ray muon data. The display interface includes GPS time and position, the detector internal clock, the measured muon flux from coincident events, and the signal frequency from each plastic scintillator module. On the right side of the screen, the amplitude of muon signals is presented as a histogram. During the initialization, the detector automatically calibrates the threshold and maps the voltage range 0–5 V to the ADC values in 0–1023 for muon signals in data acquisition and statistical analysis, ultimately generating a histogram of the amplitude distribution of cosmic muons. Due to Arduino's inability to fully sample waveforms, it can only perform a single readout of analog signals. Therefore, in order to avoid repetitive techniques caused by continuously reading the same event, a logical approach is to first read the voltage drop back to the baseline before starting the next event acquisition.

\begin{figure}[htbp]
\centering
\includegraphics[width=1\hsize]{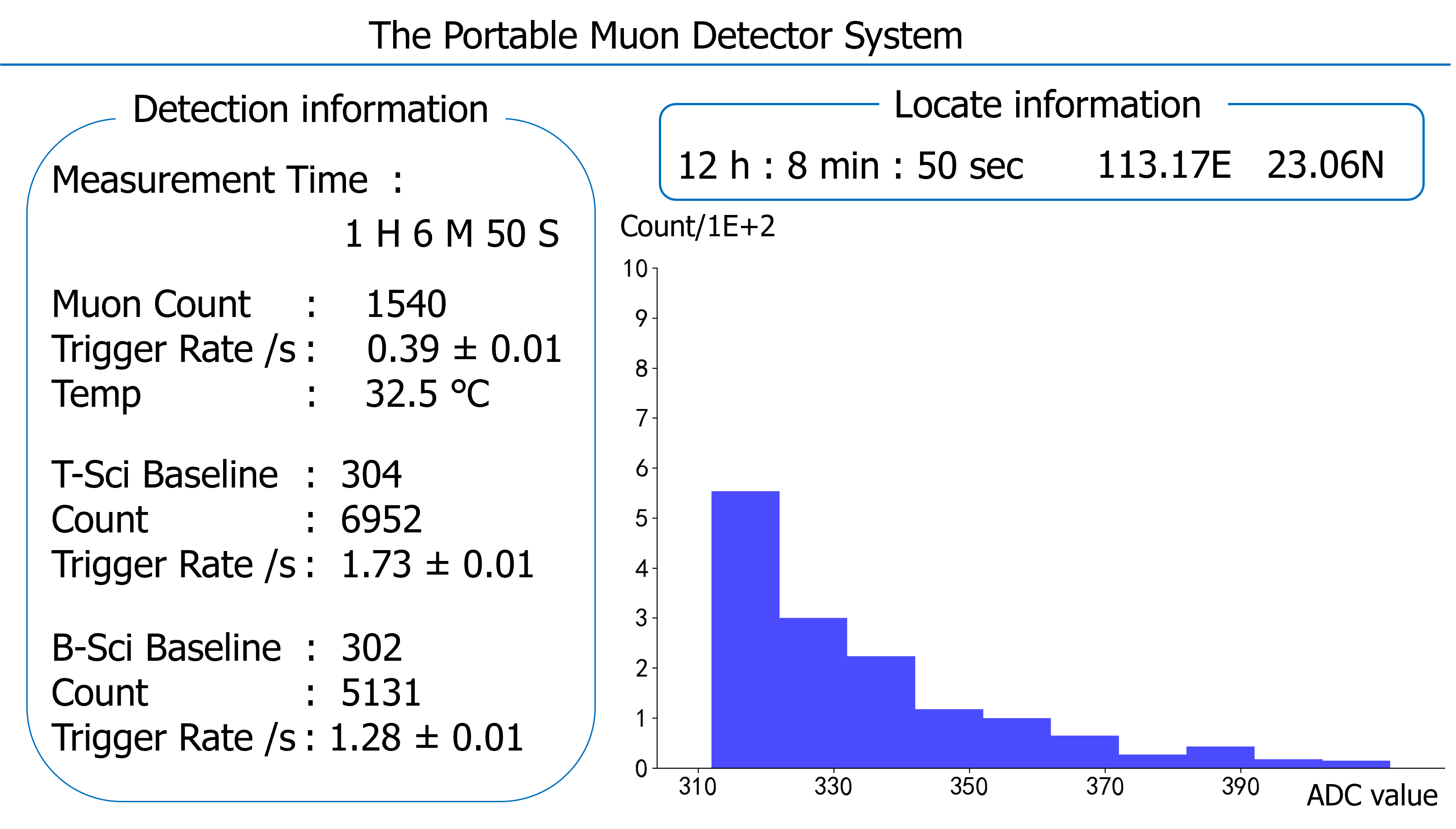}
\caption{\label{screen}The information on screen includes the detected cosmic-ray muon count and trigger rate, as well as the signal amplitude spectrum. This image provides the style of screen display information with the content in Chinese when the detector is turned on.}
\end{figure}

Whenever a cosmic-ray muon is detected, the LED on the front panel will flash, and the corresponding data produced by the analysis program are saved as a raw text file on the SD card. Each entry in the file includes the event timestamp, signal amplitude, and flux. Additionally, the system supports real-time event readout or data transfer via the serial interface, while the content exported to the computer is the same as the file on the SD card.

\begin{figure}[htbp]
\centering
\subfigure[\label{flux}]{
    \includegraphics[width=0.55\textwidth]{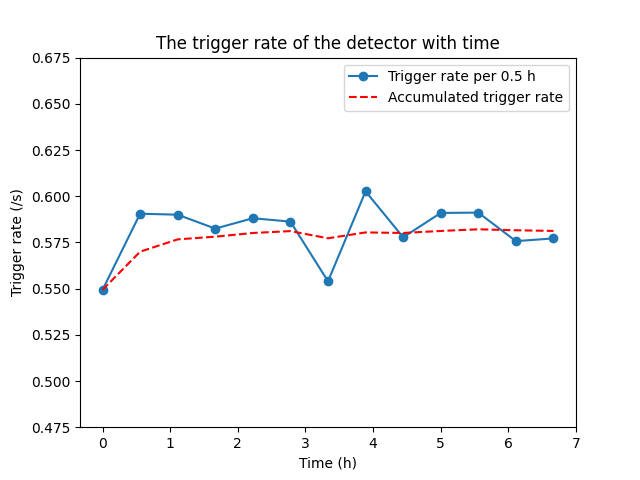}}
\quad
\subfigure[\label{histgram}]{
    \includegraphics[width=0.5\textwidth]{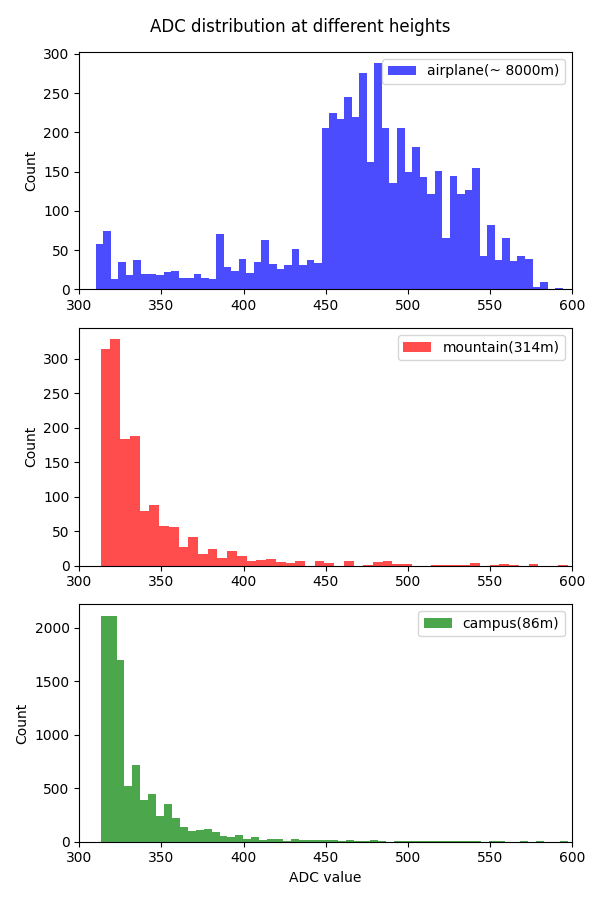}}
\caption{The portable detector maintains stable operation in several hours~(a). As the height increases, the average amplitude of the signal steps up~(b). It can be inferred that the momentum of the corresponding muons increases with the altitude.
Meanwhile, the detectors on the airplane are affected by secondary particles such as ${\pi}$/K mesons.}
\label{data}
\end{figure}

We conducted tests using the portable cosmic-ray muon detector in the first floor laboratory. It can be seen from Fig.~\ref{flux} that the detector maintained stable data collection over a relatively long period of time.
The detection efficiency is influenced by several factors: First, the circuitry in one of the coincident detector modules inevitably leads to a tunable threshold to reduce noise based on signal amplitude. The threshold will cause a loss of small signals from low-energy muons so that the detector efficiency ramps down. Second, since data processing is integrated into the Arduino platform, other functions such as read-write operation on the SD card will increase the dead time and interrupt signal acquisition performance. Moreover, the gap between two plastic scintillator detector modules will create a geometric acceptance angle. The acceptance improves the measurement accuracy, but it further reduces the effective area of the detector.

To further evaluate the detector performance, measurements were made at various altitudes, including the campus site (85 m), Liupian Mountain in Guangzhou (314 m), and a commercial airplane (${\sim}$ 8000 m) on a business trip. In general, the central value of the cosmic muon signal amplitude increased with altitude. According to the Bethe-Bloch formula~\cite{ParticleDataGroup:2024cfk}, the energy loss of cosmic muons passing through the scintillator is proportional to their kinetic energy, reflecting the fundamental relationship between the average energy of cosmic-ray muons and altitude.

\section{Array design and deployment}

\begin{figure*}[htbp]
\centering
\includegraphics[width=1\hsize]{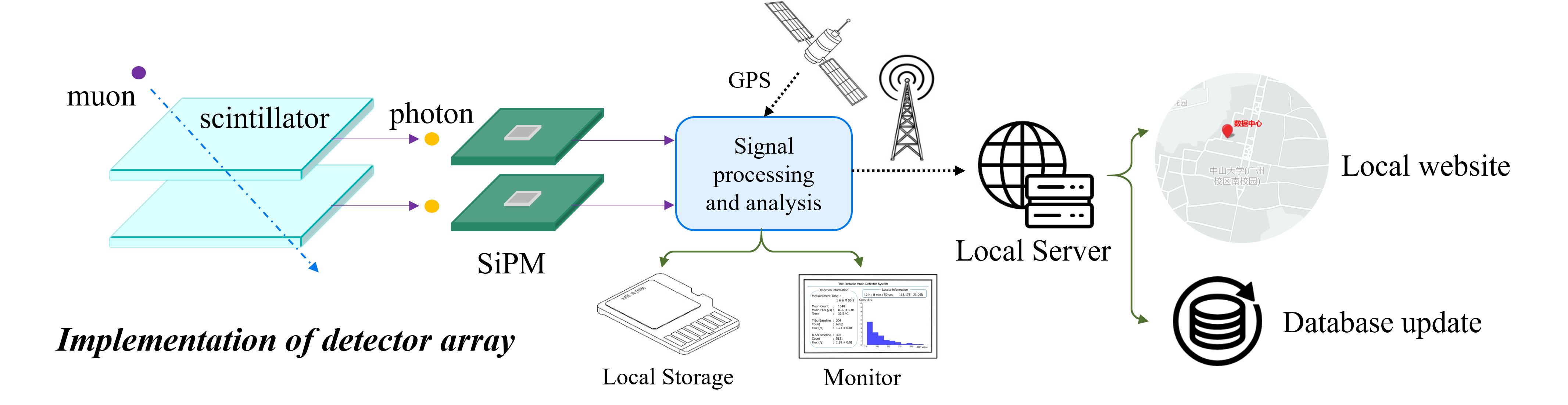}
\caption{\label{system}The detector system includes two layers of scintillators coupled with SiPMs. The detector uses GPS for timing and positioning, and the statistical results of detected cosmic-ray muons are displayed as histograms on the serial touchscreen. The collected data can be stored either on the device's SD card or transmitted to a local terminal via mobile base stations. Data are then stored in a database or used to provide information for local websites and Graphic User Interfaces (GUIs)}.
\end{figure*}

The key challenge in deploying a detector array lies in long-distance data collection and distribution. To address this issue, we implemented the widely used cellular communication technology in Internet of Things (IoT) applications, leveraging domestic mobile base stations to transmit detector data to the local server. 

\subsection{Network construction}

\begin{figure}[htbp]
\centering
\includegraphics[width=0.9\hsize]{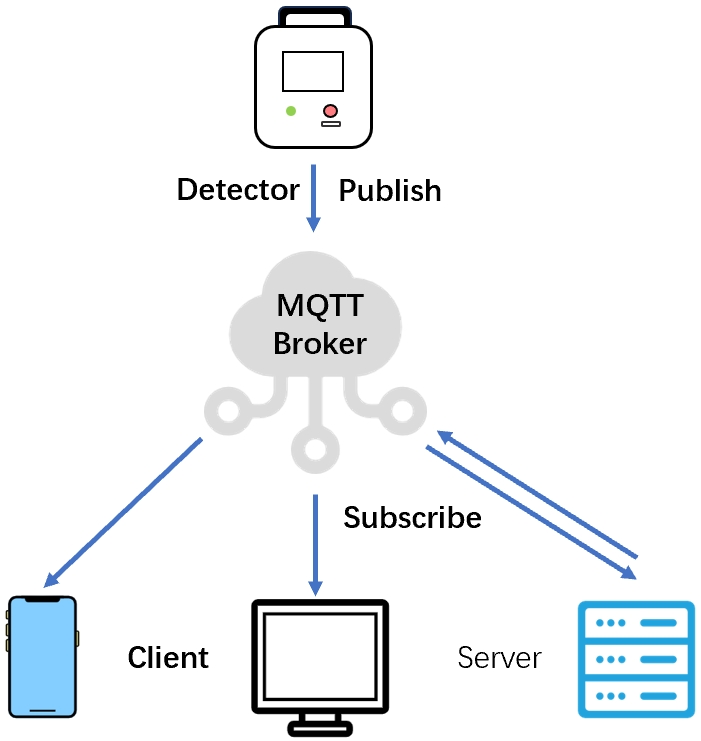}
\caption{\label{MQTT}MQTT Broker acts as an intermediary to collect and forward data from detectors, allowing detector information to be received and stored by various terminals in the form of subscriptions. }
\end{figure}

Currently, the widely used IoT technologies include Long Range (LoRa), Zigbee, and Narrowband Internet of Things (NB-IoT). LoRa operates in an unlicensed spectrum, offering a coverage ranging from 2 km to 5 km, making it suitable for privately deployed IoT networks~\cite{sinha2017survey}. Zigbee, operating on the 2.4 GHz band, provides a data rate of 250 kbps and is highly effective within a range of 100 meters~\cite{ramya2011study}. In contrast, NB-IoT is based on cellular communication technology and benefits from extensive domestic mobile base station infrastructure and long-term evolution (LTE) network deployment. As a mature and operator-deployed technology, NB-IoT offers broader coverage and deeper penetration, with a range exceeding 10 km. With a data rate of 20–250 kbps, NB-IoT is well-suited to handle the data volume generated by cosmic-ray muons, as the muon signal acquisition rate in the detector is limited by the size of the scintillator and muon flux. Furthermore, as a derivative technology of LTE, NB-IoT inherits the security mechanisms of LTE. LTE encryption ensures robust data security~\cite{hassan2020nb,martinez2019exploring}.

With the integration of NB-IoT modules, such as BC26 or MC661, we can use nationwide operator-deployed base stations to connect multiple detectors into a unified network for data transmission. In the local server setup, the Message Queuing Telemetry Transport (MQTT) protocol is employed to collect and distribute detector data.

The MQTT protocol is a lightweight application-layer communication protocol designed for message transmission between devices and servers. It is well suited for IoT environments with low bandwidth and unstable network conditions. Based on MQTT, data collected by a detector can be published to the cloud and consumed by servers or other terminals. Using the cloud as an intermediary, the system can facilitate synchronized data collection from multiple detectors~\cite{soni2017survey}. 

\begin{figure*}[!htbp]
\centering
\includegraphics[width=1\hsize]{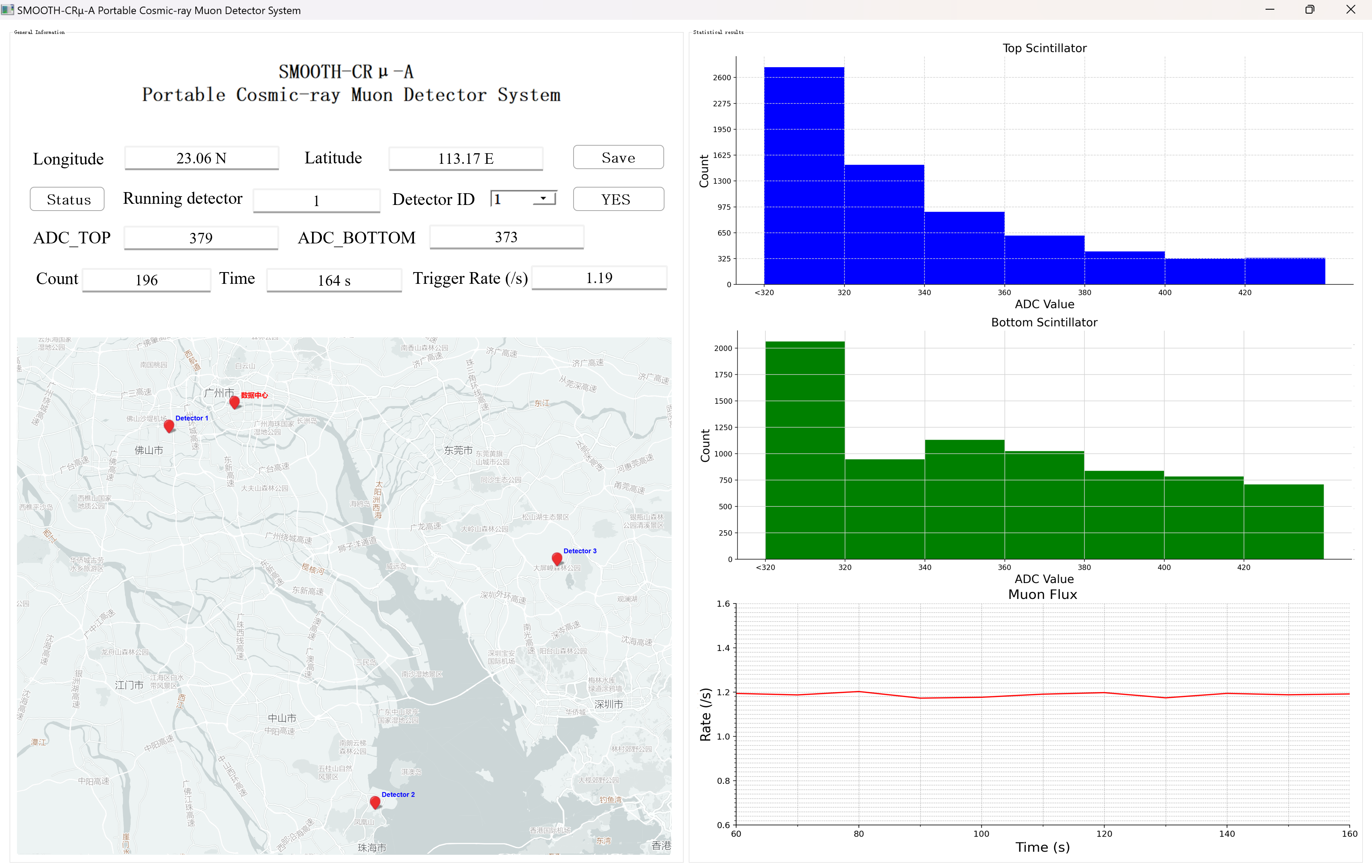}
\caption{\label{UI}The GUI program receives the real-time detector data and shows the historical data from the local database. The marker on the map provides click events by selecting the detector ID. In reality, in order to match the user requirements, a Chinese interface is designed.}
\end{figure*}

The portable cosmic-ray muon detector utilizes commercial cloud services from Alibaba cloud as an intermediary~\cite{wang2022data}. Muon data are published to the cloud in JSON format and received in real time through a subscription model for further processing. The MQTT protocol has a transmission verification function in information consumption, which can not only enable the same information to be received and viewed by multiple terminals, but also ensure that the information is fully received and prevents information loss. If the detector messages published by the cloud are not consumed promptly by terminals due to high workload or other reasons, they will be sent later when the terminals become available. This makes it possible to deliver complete messages but may result in out-of-order sequences in the message list.

\subsection{Local server deployment}

As a data consumption terminal, the system uses customized programs and access keys to receive data from the cloud. The program is deployed on a local server, where cloud data is collected, categorized, and stored, completing the construction of a remote data reception network. 

The local database is developed based on SQLite, serving as a lightweight, embedded relational database management system for efficient data storage and retrieval. Based on the identification information contained in the captured detector messages, it is possible to store the results of different detectors in separate lists.

As part of the networked construction with a benefit from the deployment of the local Apache server~\cite{laurie2003apache}, we can use Apache HTTP Server to create a visual map interface for the detector array. By using the Flask framework in the Python environment, it is possible to extend the functionality of HTML web pages, including data interaction with the local SQlite database, as well as establishing detector punctuation and click events on the map. Python-based graphical user interface (GUI) has also been developed, which mainly includes real-time monitoring of the current status of detectors and retrieval of historical data in the database. The map summaries with detector location information embedded in the GUI, while the markers on map provide clicked events by choosing one of detectors to monitor. The monitored detector id can also be chosen by the drop-down list in the GUI.

When the GUI is launched, it reads historical muon event data from the local database and performs statistical analysis to obtain a signal amplitude spectrum over a longer detection period. However, directly reading and processing large volumes of data from the database is highly inefficient, as the time cost increases with data accumulation and restricts the GUI to local network operation. To enable nationwide operation of the GUI, the Flask framework can be utilized to provide web uniform resource locator (URL) services for data transmission. By inputting the statistical results into the URL, the GUI can call the statistical information directly at the start to reduce unnecessary expenses.

For the detected cosmic-ray muon events, the information such as the amplitude and timestamp is transmitted via the cloud and displayed in the GUI, which is the same as the data saved in the SD card. In addition, these new events will accumulate continuously in the histogram. Therefore, with the displayed GUI, it is possible to monitor the detector arrays cross-linked with each other in the network and show the content on the detector screen in the form of a monitoring interface.

\subsection{Portable detector array deployment}

\begin{figure}[htbp]
\centering
\includegraphics[width=1\hsize]{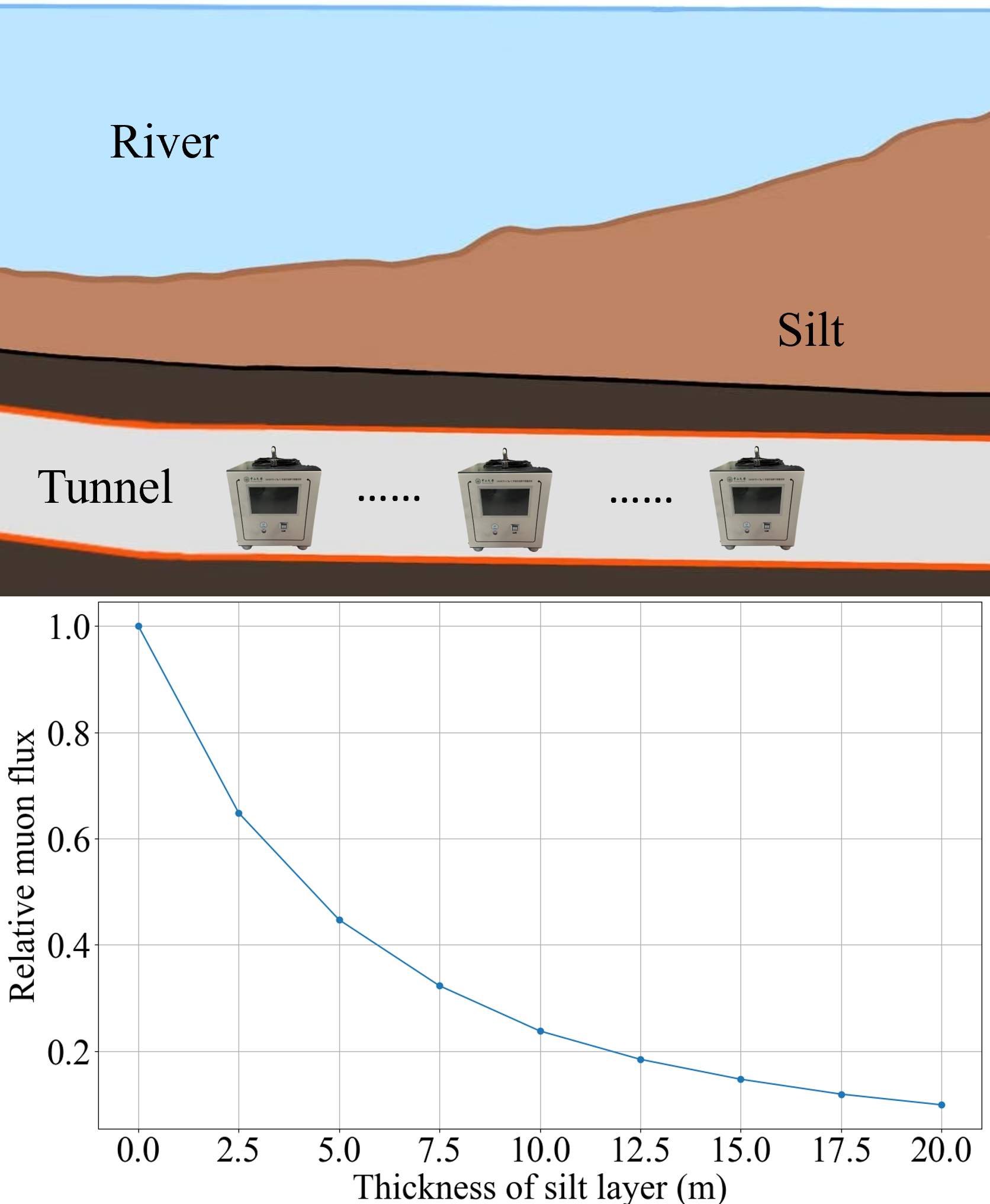}
\caption{\label{conception}A deployment of multiple IoT enabled detectors around the target object to achieve long-term monitoring of changes in the mass and thickness of large volume objects, such as monitoring silt accumulation in the traffic tunnel across the river. Here presents the results simulated by PUMAS~\cite{Niess:2022nel}, where the muon flux will vary significantly with the thickness of silt.}
\end{figure}

As shown in Fig.~\ref{conception}, it is possible to achieve long-term monitoring of changes in the mass and thickness of large volume objects by deploying multiple IoT enabled detectors around the target object, such as monitoring sediment accumulation in the traffic tunnel across the river. 

In water conservancy engineering, the long-term silt accumulation increases vertical loading on sink pipelines, which can cause shear force and excessive pressure on tunnel ceilings. As a result, the disaster will occur with the structural deformation or the broken tunnel pipe. Moreover, the uneven silt deposition creates differential settlement, generating localized bending stress concentrations in pipelines that can similarly damage the whole structure~\cite{thomas2004review}. Therefore, it is essential to monitor changes in the pattern of silt accumulation. The muography technology based on portable muon detector arrays provides a novel solution to tackle such a challenge.

With the results of PUMAS simulation~\cite{Niess:2022nel}, we can see that the flux of cosmic-ray muons will vary significantly with the thickness of the silt. As soon as the portable detector array is ready, the field test will be executed. We anticipate that the current strategy will help to avoid potential danger.

\section{Summary and outlook}

We have developed a portable cosmic-ray muon detector based on plastic scintillators and SiPMs, which has the characteristics of a complete system with low power consumption. The independent display system and data storage function ensure the long-term monitoring of cosmic-ray muons in wild field applications.

By integrating IoT technology, the data taken by a detector can be remotely collected and stored using mobile networks in areas with a coverage of base stations. Combined with the development of a local database and GUI, it is easy to conduct remote monitoring of detector status and events, laying the foundation for the construction of a full-scale detector array.

Inspired by conventional scintillator telescope arrays, the current strategy proposes to construct an ultra-wide cosmic-ray muon detector array based on the domestic infrastructure. As a network-enabled portable detector, it can be extended for applications such as educational outreach activities and the creation of cosmic-ray muon maps.

The IoT-based remote data transmission technology also has broad application prospects, including in muography, but it still faces limitations. For reconstruction of muon events, information such as the hit time, the channel number, and the signal amplitude is required. As the detector precision improves and the coverage areas expand, the data volume consumed per event transmission also increases. Therefore, this technology is currently more suitable for detectors with relatively simple structures and low data volume.

Nevertheless, IoT technology should be integrated into muography for its flexibility. With the deployment of several simply constructed detectors, pre-scanning a region can be achieved to identify potential density variations in the target object. High-precision detectors can further be used for detailed scanning, significantly reducing the resource consumption compared to full-scale measurements of large objects~\cite{Yang2022}. In addition, the detector array offers excellent stability and remote monitoring capabilities, making it well-suited for long-term observation tasks. It is expected to monitor sediment accumulation in river channels or make long-term observations of density changes in the layers of mine rock.

\begin{acknowledgments}
This project is supported by Fundamental Research Funds for the Central Universities (23xkjc017) in Sun Yat-sen University, Guangdong Basic and Applied Basic Research Foundation under Grant No. 2025A1515010669, Natural Science Foundation of Guangzhou under Grant No. 2024A04J6243 and Special Funds for the Cultivation of Guangdong College Students' Scientific and Technological Innovation under Grant No. pdjh2024a010. We appreciate the strong support from SYSU Experimental Physics Center.

\end{acknowledgments}

\section*{Data Availability Statement}
All datasets of this study are available from the corresponding author upon reasonable request.

\nocite{*}
\section*{References}
\bibliography{reference}

\begin{thebibliography}{36}%
\makeatletter
\providecommand \@ifxundefined [1]{%
 \@ifx{#1\undefined}
}%
\providecommand \@ifnum [1]{%
 \ifnum #1\expandafter \@firstoftwo
 \else \expandafter \@secondoftwo
 \fi
}%
\providecommand \@ifx [1]{%
 \ifx #1\expandafter \@firstoftwo
 \else \expandafter \@secondoftwo
 \fi
}%
\providecommand \natexlab [1]{#1}%
\providecommand \enquote  [1]{``#1''}%
\providecommand \bibnamefont  [1]{#1}%
\providecommand \bibfnamefont [1]{#1}%
\providecommand \citenamefont [1]{#1}%
\providecommand \href@noop [0]{\@secondoftwo}%
\providecommand \href [0]{\begingroup \@sanitize@url \@href}%
\providecommand \@href[1]{\@@startlink{#1}\@@href}%
\providecommand \@@href[1]{\endgroup#1\@@endlink}%
\providecommand \@sanitize@url [0]{\catcode `\\12\catcode `\$12\catcode `\&12\catcode `\#12\catcode `\^12\catcode `\_12\catcode `\%12\relax}%
\providecommand \@@startlink[1]{}%
\providecommand \@@endlink[0]{}%
\providecommand \url  [0]{\begingroup\@sanitize@url \@url }%
\providecommand \@url [1]{\endgroup\@href {#1}{\urlprefix }}%
\providecommand \urlprefix  [0]{URL }%
\providecommand \Eprint [0]{\href }%
\providecommand \doibase [0]{http://dx.doi.org/}%
\providecommand \selectlanguage [0]{\@gobble}%
\providecommand \bibinfo  [0]{\@secondoftwo}%
\providecommand \bibfield  [0]{\@secondoftwo}%
\providecommand \translation [1]{[#1]}%
\providecommand \BibitemOpen [0]{}%
\providecommand \bibitemStop [0]{}%
\providecommand \bibitemNoStop [0]{.\EOS\space}%
\providecommand \EOS [0]{\spacefactor3000\relax}%
\providecommand \BibitemShut  [1]{\csname bibitem#1\endcsname}%
\let\auto@bib@innerbib\@empty
\bibitem [{\citenamefont {Bonomi}\ \emph {et~al.}(2020)\citenamefont {Bonomi}, \citenamefont {Checchia}, \citenamefont {D'Errico}, \citenamefont {Pagano},\ and\ \citenamefont {Saracino}}]{Bonomi:2020dmm}%
  \BibitemOpen
  \bibfield  {author} {\bibinfo {author} {\bibfnamefont {G.}~\bibnamefont {Bonomi}}, \bibinfo {author} {\bibfnamefont {P.}~\bibnamefont {Checchia}}, \bibinfo {author} {\bibfnamefont {M.}~\bibnamefont {D'Errico}}, \bibinfo {author} {\bibfnamefont {D.}~\bibnamefont {Pagano}}, \ and\ \bibinfo {author} {\bibfnamefont {G.}~\bibnamefont {Saracino}},\ }\bibfield  {title} {\enquote {\bibinfo {title} {{Applications of cosmic-ray muons}},}\ }\href {\doibase 10.1016/j.ppnp.2020.103768} {\bibfield  {journal} {\bibinfo  {journal} {Prog. Part. Nucl. Phys.}\ }\textbf {\bibinfo {volume} {112}},\ \bibinfo {pages} {103768} (\bibinfo {year} {2020})}\BibitemShut {NoStop}%
\bibitem [{\citenamefont {Tanaka}\ \emph {et~al.}(2023)\citenamefont {Tanaka}, \citenamefont {Bozza}, \citenamefont {Bross}, \citenamefont {Cantoni}, \citenamefont {Catalano}, \citenamefont {Cerretto}, \citenamefont {Giammanco}, \citenamefont {Gluyas}, \citenamefont {Gnesi}, \citenamefont {Holma} \emph {et~al.}}]{tanaka2023muography}%
  \BibitemOpen
  \bibfield  {author} {\bibinfo {author} {\bibfnamefont {H.~K.}\ \bibnamefont {Tanaka}}, \bibinfo {author} {\bibfnamefont {C.}~\bibnamefont {Bozza}}, \bibinfo {author} {\bibfnamefont {A.}~\bibnamefont {Bross}}, \bibinfo {author} {\bibfnamefont {E.}~\bibnamefont {Cantoni}}, \bibinfo {author} {\bibfnamefont {O.}~\bibnamefont {Catalano}}, \bibinfo {author} {\bibfnamefont {G.}~\bibnamefont {Cerretto}}, \bibinfo {author} {\bibfnamefont {A.}~\bibnamefont {Giammanco}}, \bibinfo {author} {\bibfnamefont {J.}~\bibnamefont {Gluyas}}, \bibinfo {author} {\bibfnamefont {I.}~\bibnamefont {Gnesi}}, \bibinfo {author} {\bibfnamefont {M.}~\bibnamefont {Holma}},  \emph {et~al.},\ }\bibfield  {title} {\enquote {\bibinfo {title} {Muography},}\ }\href@noop {} {\bibfield  {journal} {\bibinfo  {journal} {Nature Reviews Methods Primers}\ }\textbf {\bibinfo {volume} {3}},\ \bibinfo {pages} {88} (\bibinfo {year} {2023})}\BibitemShut {NoStop}%
\bibitem [{\citenamefont {Bonechi}, \citenamefont {D’Alessandro},\ and\ \citenamefont {Giammanco}(2020)}]{bonechi2020atmospheric}%
  \BibitemOpen
  \bibfield  {author} {\bibinfo {author} {\bibfnamefont {L.}~\bibnamefont {Bonechi}}, \bibinfo {author} {\bibfnamefont {R.}~\bibnamefont {D’Alessandro}}, \ and\ \bibinfo {author} {\bibfnamefont {A.}~\bibnamefont {Giammanco}},\ }\bibfield  {title} {\enquote {\bibinfo {title} {Atmospheric muons as an imaging tool},}\ }\href@noop {} {\bibfield  {journal} {\bibinfo  {journal} {Reviews in Physics}\ }\textbf {\bibinfo {volume} {5}},\ \bibinfo {pages} {100038} (\bibinfo {year} {2020})}\BibitemShut {NoStop}%
\bibitem [{\citenamefont {Kaiser}(2019)}]{kaiser2019muography}%
  \BibitemOpen
  \bibfield  {author} {\bibinfo {author} {\bibfnamefont {R.}~\bibnamefont {Kaiser}},\ }\bibfield  {title} {\enquote {\bibinfo {title} {Muography: overview and future directions},}\ }\href@noop {} {\bibfield  {journal} {\bibinfo  {journal} {Philosophical Transactions of the Royal Society A}\ }\textbf {\bibinfo {volume} {377}},\ \bibinfo {pages} {20180049} (\bibinfo {year} {2019})}\BibitemShut {NoStop}%
\bibitem [{\citenamefont {Alvarez}\ \emph {et~al.}(1970)\citenamefont {Alvarez}, \citenamefont {Anderson}, \citenamefont {Bedwei}, \citenamefont {Burkhard}, \citenamefont {Fakhry}, \citenamefont {Girgis}, \citenamefont {Goneid}, \citenamefont {Hassan}, \citenamefont {Iverson}, \citenamefont {Lynch} \emph {et~al.}}]{alvarez1970search}%
  \BibitemOpen
  \bibfield  {author} {\bibinfo {author} {\bibfnamefont {L.~W.}\ \bibnamefont {Alvarez}}, \bibinfo {author} {\bibfnamefont {J.~A.}\ \bibnamefont {Anderson}}, \bibinfo {author} {\bibfnamefont {F.~E.}\ \bibnamefont {Bedwei}}, \bibinfo {author} {\bibfnamefont {J.}~\bibnamefont {Burkhard}}, \bibinfo {author} {\bibfnamefont {A.}~\bibnamefont {Fakhry}}, \bibinfo {author} {\bibfnamefont {A.}~\bibnamefont {Girgis}}, \bibinfo {author} {\bibfnamefont {A.}~\bibnamefont {Goneid}}, \bibinfo {author} {\bibfnamefont {F.}~\bibnamefont {Hassan}}, \bibinfo {author} {\bibfnamefont {D.}~\bibnamefont {Iverson}}, \bibinfo {author} {\bibfnamefont {G.}~\bibnamefont {Lynch}},  \emph {et~al.},\ }\bibfield  {title} {\enquote {\bibinfo {title} {Search for hidden chambers in the pyramids: The structure of the second pyramid of giza is determined by cosmic-ray absorption.}}\ }\href@noop {} {\bibfield  {journal} {\bibinfo  {journal} {Science}\ }\textbf {\bibinfo {volume} {167}},\ \bibinfo {pages} {832--839} (\bibinfo {year}
  {1970})}\BibitemShut {NoStop}%
\bibitem [{\citenamefont {Morishima}(2023)}]{morishima2023muography}%
  \BibitemOpen
  \bibfield  {author} {\bibinfo {author} {\bibfnamefont {K.}~\bibnamefont {Morishima}},\ }\bibfield  {title} {\enquote {\bibinfo {title} {Muography and archaeology},}\ }in\ \href@noop {} {\emph {\bibinfo {booktitle} {Cosmic Ray Muography}}}\ (\bibinfo  {publisher} {World Scientific},\ \bibinfo {year} {2023})\ pp.\ \bibinfo {pages} {233--252}\BibitemShut {NoStop}%
\bibitem [{\citenamefont {D’Errico}(2020)}]{d2020muography}%
  \BibitemOpen
  \bibfield  {author} {\bibinfo {author} {\bibfnamefont {M.}~\bibnamefont {D’Errico}},\ }\bibfield  {title} {\enquote {\bibinfo {title} {Muography applied to archaeology: Search and 3d reconstruction of hidden cavities},}\ }\href@noop {} {\bibfield  {journal} {\bibinfo  {journal} {IL Nuovo Cimento}\ }\textbf {\bibinfo {volume} {100}},\ \bibinfo {pages} {43} (\bibinfo {year} {2020})}\BibitemShut {NoStop}%
\bibitem [{\citenamefont {Tioukov}\ \emph {et~al.}(2019)\citenamefont {Tioukov}, \citenamefont {Alexandrov}, \citenamefont {Bozza}, \citenamefont {Consiglio}, \citenamefont {D’Ambrosio}, \citenamefont {De~Lellis}, \citenamefont {De~Sio}, \citenamefont {Giudicepietro}, \citenamefont {Macedonio}, \citenamefont {Miyamoto} \emph {et~al.}}]{tioukov2019first}%
  \BibitemOpen
  \bibfield  {author} {\bibinfo {author} {\bibfnamefont {V.}~\bibnamefont {Tioukov}}, \bibinfo {author} {\bibfnamefont {A.}~\bibnamefont {Alexandrov}}, \bibinfo {author} {\bibfnamefont {C.}~\bibnamefont {Bozza}}, \bibinfo {author} {\bibfnamefont {L.}~\bibnamefont {Consiglio}}, \bibinfo {author} {\bibfnamefont {N.}~\bibnamefont {D’Ambrosio}}, \bibinfo {author} {\bibfnamefont {G.}~\bibnamefont {De~Lellis}}, \bibinfo {author} {\bibfnamefont {C.}~\bibnamefont {De~Sio}}, \bibinfo {author} {\bibfnamefont {F.}~\bibnamefont {Giudicepietro}}, \bibinfo {author} {\bibfnamefont {G.}~\bibnamefont {Macedonio}}, \bibinfo {author} {\bibfnamefont {S.}~\bibnamefont {Miyamoto}},  \emph {et~al.},\ }\bibfield  {title} {\enquote {\bibinfo {title} {First muography of stromboli volcano},}\ }\href@noop {} {\bibfield  {journal} {\bibinfo  {journal} {Scientific reports}\ }\textbf {\bibinfo {volume} {9}},\ \bibinfo {pages} {6695} (\bibinfo {year} {2019})}\BibitemShut {NoStop}%
\bibitem [{\citenamefont {Riggi}\ \emph {et~al.}(2018)\citenamefont {Riggi}, \citenamefont {Antonuccio}, \citenamefont {Bandieramonte}, \citenamefont {Becciani}, \citenamefont {Bonanno}, \citenamefont {Bonanno}, \citenamefont {Bongiovanni}, \citenamefont {Fallica}, \citenamefont {Gallo}, \citenamefont {Garozzo}, \citenamefont {Grillo}, \citenamefont {La Rocca}, \citenamefont {Leonora}, \citenamefont {Longhitano}, \citenamefont {Lo Presti}, \citenamefont {Marano}, \citenamefont {Randazzo}, \citenamefont {Parasole}, \citenamefont {Petta}, \citenamefont {Riggi}, \citenamefont {Romeo}, \citenamefont {Romeo}, \citenamefont {Russo}, \citenamefont {Santagati}, \citenamefont {Timpanaro},\ and\ \citenamefont {Valvo}}]{RIGGI201816}%
  \BibitemOpen
  \bibfield  {author} {\bibinfo {author} {\bibfnamefont {F.}~\bibnamefont {Riggi}}, \bibinfo {author} {\bibfnamefont {V.}~\bibnamefont {Antonuccio}}, \bibinfo {author} {\bibfnamefont {M.}~\bibnamefont {Bandieramonte}}, \bibinfo {author} {\bibfnamefont {U.}~\bibnamefont {Becciani}}, \bibinfo {author} {\bibfnamefont {G.}~\bibnamefont {Bonanno}}, \bibinfo {author} {\bibfnamefont {D.}~\bibnamefont {Bonanno}}, \bibinfo {author} {\bibfnamefont {D.}~\bibnamefont {Bongiovanni}}, \bibinfo {author} {\bibfnamefont {P.}~\bibnamefont {Fallica}}, \bibinfo {author} {\bibfnamefont {G.}~\bibnamefont {Gallo}}, \bibinfo {author} {\bibfnamefont {S.}~\bibnamefont {Garozzo}}, \bibinfo {author} {\bibfnamefont {A.}~\bibnamefont {Grillo}}, \bibinfo {author} {\bibfnamefont {P.}~\bibnamefont {La Rocca}}, \bibinfo {author} {\bibfnamefont {E.}~\bibnamefont {Leonora}}, \bibinfo {author} {\bibfnamefont {F.}~\bibnamefont {Longhitano}}, \bibinfo {author} {\bibfnamefont {D.}~\bibnamefont {Lo Presti}}, \bibinfo {author} {\bibfnamefont
  {D.}~\bibnamefont {Marano}}, \bibinfo {author} {\bibfnamefont {N.}~\bibnamefont {Randazzo}}, \bibinfo {author} {\bibfnamefont {O.}~\bibnamefont {Parasole}}, \bibinfo {author} {\bibfnamefont {C.}~\bibnamefont {Petta}}, \bibinfo {author} {\bibfnamefont {S.}~\bibnamefont {Riggi}}, \bibinfo {author} {\bibfnamefont {G.}~\bibnamefont {Romeo}}, \bibinfo {author} {\bibfnamefont {M.}~\bibnamefont {Romeo}}, \bibinfo {author} {\bibfnamefont {G.}~\bibnamefont {Russo}}, \bibinfo {author} {\bibfnamefont {G.}~\bibnamefont {Santagati}}, \bibinfo {author} {\bibfnamefont {M.}~\bibnamefont {Timpanaro}}, \ and\ \bibinfo {author} {\bibfnamefont {G.}~\bibnamefont {Valvo}},\ }\bibfield  {title} {\enquote {\bibinfo {title} {The muon portal project: Commissioning of the full detector and first results},}\ }\href {\doibase https://doi.org/10.1016/j.nima.2017.10.006} {\bibfield  {journal} {\bibinfo  {journal} {Nuclear Instruments and Methods in Physics Research Section A: Accelerators, Spectrometers, Detectors and Associated
  Equipment}\ }\textbf {\bibinfo {volume} {912}},\ \bibinfo {pages} {16--19} (\bibinfo {year} {2018})},\ \bibinfo {note} {new Developments In Photodetection 2017}\BibitemShut {NoStop}%
\bibitem [{\citenamefont {Anghel}\ \emph {et~al.}(2012)\citenamefont {Anghel}, \citenamefont {Armitage}, \citenamefont {Botte}, \citenamefont {Boudjemline}, \citenamefont {Bryman}, \citenamefont {Bueno}, \citenamefont {Charles}, \citenamefont {Cousins}, \citenamefont {Erlandson}, \citenamefont {Gallant}, \citenamefont {Gazit}, \citenamefont {Golovko}, \citenamefont {Hydomako}, \citenamefont {Jewett}, \citenamefont {Jonkmans}, \citenamefont {Liu}, \citenamefont {Magill}, \citenamefont {Noel}, \citenamefont {Oakham}, \citenamefont {Robichaud}, \citenamefont {Stocki}, \citenamefont {Thompson},\ and\ \citenamefont {Waller}}]{6551201}%
  \BibitemOpen
  \bibfield  {author} {\bibinfo {author} {\bibfnamefont {V.}~\bibnamefont {Anghel}}, \bibinfo {author} {\bibfnamefont {J.}~\bibnamefont {Armitage}}, \bibinfo {author} {\bibfnamefont {J.}~\bibnamefont {Botte}}, \bibinfo {author} {\bibfnamefont {K.}~\bibnamefont {Boudjemline}}, \bibinfo {author} {\bibfnamefont {D.}~\bibnamefont {Bryman}}, \bibinfo {author} {\bibfnamefont {J.}~\bibnamefont {Bueno}}, \bibinfo {author} {\bibfnamefont {E.}~\bibnamefont {Charles}}, \bibinfo {author} {\bibfnamefont {T.}~\bibnamefont {Cousins}}, \bibinfo {author} {\bibfnamefont {A.}~\bibnamefont {Erlandson}}, \bibinfo {author} {\bibfnamefont {G.}~\bibnamefont {Gallant}}, \bibinfo {author} {\bibfnamefont {R.}~\bibnamefont {Gazit}}, \bibinfo {author} {\bibfnamefont {V.}~\bibnamefont {Golovko}}, \bibinfo {author} {\bibfnamefont {R.}~\bibnamefont {Hydomako}}, \bibinfo {author} {\bibfnamefont {C.}~\bibnamefont {Jewett}}, \bibinfo {author} {\bibfnamefont {G.}~\bibnamefont {Jonkmans}}, \bibinfo {author} {\bibfnamefont {Z.}~\bibnamefont
  {Liu}}, \bibinfo {author} {\bibfnamefont {M.}~\bibnamefont {Magill}}, \bibinfo {author} {\bibfnamefont {S.}~\bibnamefont {Noel}}, \bibinfo {author} {\bibfnamefont {G.}~\bibnamefont {Oakham}}, \bibinfo {author} {\bibfnamefont {A.}~\bibnamefont {Robichaud}}, \bibinfo {author} {\bibfnamefont {T.}~\bibnamefont {Stocki}}, \bibinfo {author} {\bibfnamefont {M.}~\bibnamefont {Thompson}}, \ and\ \bibinfo {author} {\bibfnamefont {D.}~\bibnamefont {Waller}},\ }\bibfield  {title} {\enquote {\bibinfo {title} {Construction, commissioning and first data from the cript muon tomography project},}\ }in\ \href {\doibase 10.1109/NSSMIC.2012.6551201} {\emph {\bibinfo {booktitle} {2012 IEEE Nuclear Science Symposium and Medical Imaging Conference Record (NSS/MIC)}}}\ (\bibinfo {year} {2012})\ pp.\ \bibinfo {pages} {738--742}\BibitemShut {NoStop}%
\bibitem [{\citenamefont {Tanaka}\ \emph {et~al.}(2022)\citenamefont {Tanaka}, \citenamefont {Aichi}, \citenamefont {Balogh}, \citenamefont {Bozza}, \citenamefont {Coniglione}, \citenamefont {Gluyas}, \citenamefont {Hayashi}, \citenamefont {Holma}, \citenamefont {Joutsenvaara}, \citenamefont {Kamoshida} \emph {et~al.}}]{tanaka2022periodic}%
  \BibitemOpen
  \bibfield  {author} {\bibinfo {author} {\bibfnamefont {H.~K.}\ \bibnamefont {Tanaka}}, \bibinfo {author} {\bibfnamefont {M.}~\bibnamefont {Aichi}}, \bibinfo {author} {\bibfnamefont {S.~J.}\ \bibnamefont {Balogh}}, \bibinfo {author} {\bibfnamefont {C.}~\bibnamefont {Bozza}}, \bibinfo {author} {\bibfnamefont {R.}~\bibnamefont {Coniglione}}, \bibinfo {author} {\bibfnamefont {J.}~\bibnamefont {Gluyas}}, \bibinfo {author} {\bibfnamefont {N.}~\bibnamefont {Hayashi}}, \bibinfo {author} {\bibfnamefont {M.}~\bibnamefont {Holma}}, \bibinfo {author} {\bibfnamefont {J.}~\bibnamefont {Joutsenvaara}}, \bibinfo {author} {\bibfnamefont {O.}~\bibnamefont {Kamoshida}},  \emph {et~al.},\ }\bibfield  {title} {\enquote {\bibinfo {title} {Periodic sea-level oscillation in tokyo bay detected with the tokyo-bay seafloor hyper-kilometric submarine deep detector (ts-hkmsdd)},}\ }\href@noop {} {\bibfield  {journal} {\bibinfo  {journal} {Scientific reports}\ }\textbf {\bibinfo {volume} {12}},\ \bibinfo {pages} {6097} (\bibinfo {year}
  {2022})}\BibitemShut {NoStop}%
\bibitem [{\citenamefont {Leone}\ \emph {et~al.}(2021)\citenamefont {Leone}, \citenamefont {Tanaka}, \citenamefont {Holma}, \citenamefont {Kuusiniemi}, \citenamefont {Varga}, \citenamefont {Olah}, \citenamefont {Presti}, \citenamefont {Gallo}, \citenamefont {Monaco}, \citenamefont {Ferlito} \emph {et~al.}}]{leone2021muography}%
  \BibitemOpen
  \bibfield  {author} {\bibinfo {author} {\bibfnamefont {G.}~\bibnamefont {Leone}}, \bibinfo {author} {\bibfnamefont {H.~K.}\ \bibnamefont {Tanaka}}, \bibinfo {author} {\bibfnamefont {M.}~\bibnamefont {Holma}}, \bibinfo {author} {\bibfnamefont {P.}~\bibnamefont {Kuusiniemi}}, \bibinfo {author} {\bibfnamefont {D.}~\bibnamefont {Varga}}, \bibinfo {author} {\bibfnamefont {L.}~\bibnamefont {Olah}}, \bibinfo {author} {\bibfnamefont {D.~L.}\ \bibnamefont {Presti}}, \bibinfo {author} {\bibfnamefont {G.}~\bibnamefont {Gallo}}, \bibinfo {author} {\bibfnamefont {C.}~\bibnamefont {Monaco}}, \bibinfo {author} {\bibfnamefont {C.}~\bibnamefont {Ferlito}},  \emph {et~al.},\ }\bibfield  {title} {\enquote {\bibinfo {title} {Muography as a new complementary tool in monitoring volcanic hazard: implications for early warning systems},}\ }\href@noop {} {\bibfield  {journal} {\bibinfo  {journal} {Proceedings of the Royal Society A}\ }\textbf {\bibinfo {volume} {477}},\ \bibinfo {pages} {20210320} (\bibinfo {year}
  {2021})}\BibitemShut {NoStop}%
\bibitem [{\citenamefont {Ol{\'a}h}\ \emph {et~al.}(2018)\citenamefont {Ol{\'a}h}, \citenamefont {Tanaka}, \citenamefont {Ohminato},\ and\ \citenamefont {Varga}}]{olah2018high}%
  \BibitemOpen
  \bibfield  {author} {\bibinfo {author} {\bibfnamefont {L.}~\bibnamefont {Ol{\'a}h}}, \bibinfo {author} {\bibfnamefont {H.~K.}\ \bibnamefont {Tanaka}}, \bibinfo {author} {\bibfnamefont {T.}~\bibnamefont {Ohminato}}, \ and\ \bibinfo {author} {\bibfnamefont {D.}~\bibnamefont {Varga}},\ }\bibfield  {title} {\enquote {\bibinfo {title} {High-definition and low-noise muography of the sakurajima volcano with gaseous tracking detectors},}\ }\href@noop {} {\bibfield  {journal} {\bibinfo  {journal} {Scientific reports}\ }\textbf {\bibinfo {volume} {8}},\ \bibinfo {pages} {3207} (\bibinfo {year} {2018})}\BibitemShut {NoStop}%
\bibitem [{\citenamefont {Tanaka}(2020)}]{tanaka2020development}%
  \BibitemOpen
  \bibfield  {author} {\bibinfo {author} {\bibfnamefont {H.~K.}\ \bibnamefont {Tanaka}},\ }\bibfield  {title} {\enquote {\bibinfo {title} {Development of the muographic tephra deposit monitoring system},}\ }\href@noop {} {\bibfield  {journal} {\bibinfo  {journal} {Scientific Reports}\ }\textbf {\bibinfo {volume} {10}},\ \bibinfo {pages} {14820} (\bibinfo {year} {2020})}\BibitemShut {NoStop}%
\bibitem [{\citenamefont {Liu}\ \emph {et~al.}(2024)\citenamefont {Liu}, \citenamefont {Yao}, \citenamefont {Niu}, \citenamefont {Li}, \citenamefont {Tian}, \citenamefont {Li}, \citenamefont {Luo}, \citenamefont {Jin}, \citenamefont {Gao}, \citenamefont {Rong} \emph {et~al.}}]{liu2024deep}%
  \BibitemOpen
  \bibfield  {author} {\bibinfo {author} {\bibfnamefont {G.}~\bibnamefont {Liu}}, \bibinfo {author} {\bibfnamefont {K.}~\bibnamefont {Yao}}, \bibinfo {author} {\bibfnamefont {F.}~\bibnamefont {Niu}}, \bibinfo {author} {\bibfnamefont {Z.}~\bibnamefont {Li}}, \bibinfo {author} {\bibfnamefont {H.}~\bibnamefont {Tian}}, \bibinfo {author} {\bibfnamefont {J.}~\bibnamefont {Li}}, \bibinfo {author} {\bibfnamefont {X.}~\bibnamefont {Luo}}, \bibinfo {author} {\bibfnamefont {L.}~\bibnamefont {Jin}}, \bibinfo {author} {\bibfnamefont {J.}~\bibnamefont {Gao}}, \bibinfo {author} {\bibfnamefont {J.}~\bibnamefont {Rong}},  \emph {et~al.},\ }\bibfield  {title} {\enquote {\bibinfo {title} {Deep investigation of muography in discovering geological structures in mineral exploration: a case study of zaozigou gold mine},}\ }\href@noop {} {\bibfield  {journal} {\bibinfo  {journal} {Geophysical Journal International}\ }\textbf {\bibinfo {volume} {237}},\ \bibinfo {pages} {588--603} (\bibinfo {year} {2024})}\BibitemShut {NoStop}%
\bibitem [{\citenamefont {Amenomori}\ \emph {et~al.}(1990)\citenamefont {Amenomori}, \citenamefont {Nanjo}, \citenamefont {Hotta}, \citenamefont {Ohta}, \citenamefont {Kasahara}, \citenamefont {Saito}, \citenamefont {Yuda}, \citenamefont {Mizutani}, \citenamefont {Shirai}, \citenamefont {Tateyama} \emph {et~al.}}]{amenomori1990development}%
  \BibitemOpen
  \bibfield  {author} {\bibinfo {author} {\bibfnamefont {M.}~\bibnamefont {Amenomori}}, \bibinfo {author} {\bibfnamefont {H.}~\bibnamefont {Nanjo}}, \bibinfo {author} {\bibfnamefont {N.}~\bibnamefont {Hotta}}, \bibinfo {author} {\bibfnamefont {I.}~\bibnamefont {Ohta}}, \bibinfo {author} {\bibfnamefont {K.}~\bibnamefont {Kasahara}}, \bibinfo {author} {\bibfnamefont {T.}~\bibnamefont {Saito}}, \bibinfo {author} {\bibfnamefont {T.}~\bibnamefont {Yuda}}, \bibinfo {author} {\bibfnamefont {K.}~\bibnamefont {Mizutani}}, \bibinfo {author} {\bibfnamefont {T.}~\bibnamefont {Shirai}}, \bibinfo {author} {\bibfnamefont {N.}~\bibnamefont {Tateyama}},  \emph {et~al.},\ }\bibfield  {title} {\enquote {\bibinfo {title} {Development and performance test of a prototype air shower array for search for gamma ray point sources in the very high energy region},}\ }\href@noop {} {\bibfield  {journal} {\bibinfo  {journal} {Nuclear Instruments and Methods in Physics Research Section A: Accelerators, Spectrometers, Detectors and
  Associated Equipment}\ }\textbf {\bibinfo {volume} {288}},\ \bibinfo {pages} {619--631} (\bibinfo {year} {1990})}\BibitemShut {NoStop}%
\bibitem [{\citenamefont {Davier}(2005)}]{Davier:2005id}%
  \BibitemOpen
  \bibfield  {author} {\bibinfo {author} {\bibfnamefont {M.}~\bibnamefont {Davier}},\ }\bibfield  {title} {\enquote {\bibinfo {title} {{Multi-messenger astronomy}},}\ }\href {\doibase 10.1016/j.nuclphysbps.2005.01.136} {\bibfield  {journal} {\bibinfo  {journal} {Nucl. Phys. B Proc. Suppl.}\ }\textbf {\bibinfo {volume} {143}},\ \bibinfo {pages} {395--406} (\bibinfo {year} {2005})}\BibitemShut {NoStop}%
\bibitem [{\citenamefont {Neronov}(2019)}]{neronov2019introduction}%
  \BibitemOpen
  \bibfield  {author} {\bibinfo {author} {\bibfnamefont {A.}~\bibnamefont {Neronov}},\ }\bibfield  {title} {\enquote {\bibinfo {title} {Introduction to multi-messenger astronomy},}\ }in\ \href@noop {} {\emph {\bibinfo {booktitle} {Journal of Physics: Conference Series}}},\ Vol.\ \bibinfo {volume} {1263}\ (\bibinfo {organization} {IOP Publishing},\ \bibinfo {year} {2019})\ p.\ \bibinfo {pages} {012001}\BibitemShut {NoStop}%
\bibitem [{\citenamefont {Li}\ \emph {et~al.}(2018)\citenamefont {Li} \emph {et~al.}}]{Li:2018vpy}%
  \BibitemOpen
  \bibfield  {author} {\bibinfo {author} {\bibfnamefont {C.}~\bibnamefont {Li}} \emph {et~al.},\ }\bibfield  {title} {\enquote {\bibinfo {title} {{Measurement of muonic and electromagnetic components in cosmic ray air showers using LHAASO-KM2A prototype array}},}\ }\href {\doibase 10.1103/PhysRevD.98.042001} {\bibfield  {journal} {\bibinfo  {journal} {Phys. Rev. D}\ }\textbf {\bibinfo {volume} {98}},\ \bibinfo {pages} {042001} (\bibinfo {year} {2018})}\BibitemShut {NoStop}%
\bibitem [{\citenamefont {Cao}\ \emph {et~al.}(2025)\citenamefont {Cao} \emph {et~al.}}]{LHAASO:2024kbg}%
  \BibitemOpen
  \bibfield  {author} {\bibinfo {author} {\bibfnamefont {Z.}~\bibnamefont {Cao}} \emph {et~al.} (\bibinfo {collaboration} {LHAASO}),\ }\bibfield  {title} {\enquote {\bibinfo {title} {{Data quality control system and long-term performance monitor of LHAASO-KM2A}},}\ }\href {\doibase 10.1016/j.astropartphys.2024.103029} {\bibfield  {journal} {\bibinfo  {journal} {Astropart. Phys.}\ }\textbf {\bibinfo {volume} {164}},\ \bibinfo {pages} {103029} (\bibinfo {year} {2025})},\ \Eprint {http://arxiv.org/abs/2405.11826} {arXiv:2405.11826 [astro-ph.IM]} \BibitemShut {NoStop}%
\bibitem [{\citenamefont {Cao}\ \emph {et~al.}(2021)\citenamefont {Cao} \emph {et~al.}}]{LHAASO:2021gok}%
  \BibitemOpen
  \bibfield  {author} {\bibinfo {author} {\bibfnamefont {Z.}~\bibnamefont {Cao}} \emph {et~al.} (\bibinfo {collaboration} {LHAASO}),\ }\bibfield  {title} {\enquote {\bibinfo {title} {{Ultrahigh-energy photons up to 1.4 petaelectronvolts from 12 $\gamma$-ray Galactic sources}},}\ }\href {\doibase 10.1038/s41586-021-03498-z} {\bibfield  {journal} {\bibinfo  {journal} {Nature}\ }\textbf {\bibinfo {volume} {594}},\ \bibinfo {pages} {33--36} (\bibinfo {year} {2021})}\BibitemShut {NoStop}%
\bibitem [{\citenamefont {Cao}\ \emph {et~al.}(2023)\citenamefont {Cao} \emph {et~al.}}]{LHAASO:2023gne}%
  \BibitemOpen
  \bibfield  {author} {\bibinfo {author} {\bibfnamefont {Z.}~\bibnamefont {Cao}} \emph {et~al.} (\bibinfo {collaboration} {LHAASO}),\ }\bibfield  {title} {\enquote {\bibinfo {title} {{Measurement of Ultra-High-Energy Diffuse Gamma-Ray Emission of the Galactic Plane from 10~TeV to 1~PeV with LHAASO-KM2A}},}\ }\href {\doibase 10.1103/PhysRevLett.131.151001} {\bibfield  {journal} {\bibinfo  {journal} {Phys. Rev. Lett.}\ }\textbf {\bibinfo {volume} {131}},\ \bibinfo {pages} {151001} (\bibinfo {year} {2023})},\ \Eprint {http://arxiv.org/abs/2305.05372} {arXiv:2305.05372 [astro-ph.HE]} \BibitemShut {NoStop}%
\bibitem [{\citenamefont {Zhong}\ \emph {et~al.}(2025)\citenamefont {Zhong}, \citenamefont {Shah}, \citenamefont {Zhou},\ and\ \citenamefont {Tang}}]{zhong2025enhancing}%
  \BibitemOpen
  \bibfield  {author} {\bibinfo {author} {\bibfnamefont {J.}~\bibnamefont {Zhong}}, \bibinfo {author} {\bibfnamefont {N.~A.}\ \bibnamefont {Shah}}, \bibinfo {author} {\bibfnamefont {J.}~\bibnamefont {Zhou}}, \ and\ \bibinfo {author} {\bibfnamefont {J.}~\bibnamefont {Tang}},\ }\bibfield  {title} {\enquote {\bibinfo {title} {Enhancing plastic scintillator performance through advanced injection molding techniques},}\ }\href@noop {} {\bibfield  {journal} {\bibinfo  {journal} {Radiation Physics and Chemistry}\ }\textbf {\bibinfo {volume} {226}},\ \bibinfo {pages} {112193} (\bibinfo {year} {2025})}\BibitemShut {NoStop}%
\bibitem [{\citenamefont {Axani}(2019)}]{axani2019physics}%
  \BibitemOpen
  \bibfield  {author} {\bibinfo {author} {\bibfnamefont {S.~N.}\ \bibnamefont {Axani}},\ }\bibfield  {title} {\enquote {\bibinfo {title} {The physics behind the cosmicwatch desktop muon detectors},}\ }\href@noop {} {\bibfield  {journal} {\bibinfo  {journal} {arXiv preprint arXiv:1908.00146}\ } (\bibinfo {year} {2019})}\BibitemShut {NoStop}%
\bibitem [{\citenamefont {Navas}\ \emph {et~al.}(2024)\citenamefont {Navas} \emph {et~al.}}]{ParticleDataGroup:2024cfk}%
  \BibitemOpen
  \bibfield  {author} {\bibinfo {author} {\bibfnamefont {S.}~\bibnamefont {Navas}} \emph {et~al.} (\bibinfo {collaboration} {Particle Data Group}),\ }\bibfield  {title} {\enquote {\bibinfo {title} {{Review of particle physics}},}\ }\href {\doibase 10.1103/PhysRevD.110.030001} {\bibfield  {journal} {\bibinfo  {journal} {Phys. Rev. D}\ }\textbf {\bibinfo {volume} {110}},\ \bibinfo {pages} {030001} (\bibinfo {year} {2024})}\BibitemShut {NoStop}%
\bibitem [{\citenamefont {Sinha}, \citenamefont {Wei},\ and\ \citenamefont {Hwang}(2017)}]{sinha2017survey}%
  \BibitemOpen
  \bibfield  {author} {\bibinfo {author} {\bibfnamefont {R.~S.}\ \bibnamefont {Sinha}}, \bibinfo {author} {\bibfnamefont {Y.}~\bibnamefont {Wei}}, \ and\ \bibinfo {author} {\bibfnamefont {S.-H.}\ \bibnamefont {Hwang}},\ }\bibfield  {title} {\enquote {\bibinfo {title} {A survey on lpwa technology: Lora and nb-iot},}\ }\href@noop {} {\bibfield  {journal} {\bibinfo  {journal} {Ict Express}\ }\textbf {\bibinfo {volume} {3}},\ \bibinfo {pages} {14--21} (\bibinfo {year} {2017})}\BibitemShut {NoStop}%
\bibitem [{\citenamefont {Ramya}, \citenamefont {Shanmugaraj},\ and\ \citenamefont {Prabakaran}(2011)}]{ramya2011study}%
  \BibitemOpen
  \bibfield  {author} {\bibinfo {author} {\bibfnamefont {C.~M.}\ \bibnamefont {Ramya}}, \bibinfo {author} {\bibfnamefont {M.}~\bibnamefont {Shanmugaraj}}, \ and\ \bibinfo {author} {\bibfnamefont {R.}~\bibnamefont {Prabakaran}},\ }\bibfield  {title} {\enquote {\bibinfo {title} {Study on zigbee technology},}\ }in\ \href@noop {} {\emph {\bibinfo {booktitle} {2011 3rd international conference on electronics computer technology}}},\ Vol.~\bibinfo {volume} {6}\ (\bibinfo {organization} {IEEE},\ \bibinfo {year} {2011})\ pp.\ \bibinfo {pages} {297--301}\BibitemShut {NoStop}%
\bibitem [{\citenamefont {Hassan}\ \emph {et~al.}(2020)\citenamefont {Hassan}, \citenamefont {Ali}, \citenamefont {Mokhtar}, \citenamefont {Saeed},\ and\ \citenamefont {Chaudhari}}]{hassan2020nb}%
  \BibitemOpen
  \bibfield  {author} {\bibinfo {author} {\bibfnamefont {M.~B.}\ \bibnamefont {Hassan}}, \bibinfo {author} {\bibfnamefont {E.~S.}\ \bibnamefont {Ali}}, \bibinfo {author} {\bibfnamefont {R.~A.}\ \bibnamefont {Mokhtar}}, \bibinfo {author} {\bibfnamefont {R.~A.}\ \bibnamefont {Saeed}}, \ and\ \bibinfo {author} {\bibfnamefont {B.~S.}\ \bibnamefont {Chaudhari}},\ }\bibfield  {title} {\enquote {\bibinfo {title} {Nb-iot: Concepts, applications, and deployment challenges},}\ }in\ \href@noop {} {\emph {\bibinfo {booktitle} {LPWAN Technologies for IoT and M2M Applications}}}\ (\bibinfo  {publisher} {Elsevier},\ \bibinfo {year} {2020})\ pp.\ \bibinfo {pages} {119--144}\BibitemShut {NoStop}%
\bibitem [{\citenamefont {Martinez}\ \emph {et~al.}(2019)\citenamefont {Martinez}, \citenamefont {Adelantado}, \citenamefont {Bartoli},\ and\ \citenamefont {Vilajosana}}]{martinez2019exploring}%
  \BibitemOpen
  \bibfield  {author} {\bibinfo {author} {\bibfnamefont {B.}~\bibnamefont {Martinez}}, \bibinfo {author} {\bibfnamefont {F.}~\bibnamefont {Adelantado}}, \bibinfo {author} {\bibfnamefont {A.}~\bibnamefont {Bartoli}}, \ and\ \bibinfo {author} {\bibfnamefont {X.}~\bibnamefont {Vilajosana}},\ }\bibfield  {title} {\enquote {\bibinfo {title} {Exploring the performance boundaries of nb-iot},}\ }\href@noop {} {\bibfield  {journal} {\bibinfo  {journal} {IEEE Internet of Things Journal}\ }\textbf {\bibinfo {volume} {6}},\ \bibinfo {pages} {5702--5712} (\bibinfo {year} {2019})}\BibitemShut {NoStop}%
\bibitem [{\citenamefont {Soni}\ and\ \citenamefont {Makwana}(2017)}]{soni2017survey}%
  \BibitemOpen
  \bibfield  {author} {\bibinfo {author} {\bibfnamefont {D.}~\bibnamefont {Soni}}\ and\ \bibinfo {author} {\bibfnamefont {A.}~\bibnamefont {Makwana}},\ }\bibfield  {title} {\enquote {\bibinfo {title} {A survey on mqtt: a protocol of internet of things (iot)},}\ }in\ \href@noop {} {\emph {\bibinfo {booktitle} {International conference on telecommunication, power analysis and computing techniques (ICTPACT-2017)}}},\ Vol.~\bibinfo {volume} {20}\ (\bibinfo {year} {2017})\ pp.\ \bibinfo {pages} {173--177}\BibitemShut {NoStop}%
\bibitem [{\citenamefont {Wang}(2022)}]{wang2022data}%
  \BibitemOpen
  \bibfield  {author} {\bibinfo {author} {\bibfnamefont {L.}~\bibnamefont {Wang}},\ }\bibfield  {title} {\enquote {\bibinfo {title} {Data on cloud practice based on alibaba cloud iot platform},}\ }in\ \href@noop {} {\emph {\bibinfo {booktitle} {Smart Communications, Intelligent Algorithms and Interactive Methods: Proceedings of 4th International Conference on Wireless Communications and Applications (ICWCA 2020)}}}\ (\bibinfo {organization} {Springer},\ \bibinfo {year} {2022})\ pp.\ \bibinfo {pages} {143--147}\BibitemShut {NoStop}%
\bibitem [{\citenamefont {Laurie}\ and\ \citenamefont {Laurie}(2003)}]{laurie2003apache}%
  \BibitemOpen
  \bibfield  {author} {\bibinfo {author} {\bibfnamefont {B.}~\bibnamefont {Laurie}}\ and\ \bibinfo {author} {\bibfnamefont {P.}~\bibnamefont {Laurie}},\ }\href@noop {} {\emph {\bibinfo {title} {Apache: The definitive guide}}}\ (\bibinfo  {publisher} {" O'Reilly Media, Inc."},\ \bibinfo {year} {2003})\BibitemShut {NoStop}%
\bibitem [{\citenamefont {Niess}(2022)}]{Niess:2022nel}%
  \BibitemOpen
  \bibfield  {author} {\bibinfo {author} {\bibfnamefont {V.}~\bibnamefont {Niess}},\ }\bibfield  {title} {\enquote {\bibinfo {title} {{The PUMAS library}},}\ }\href {\doibase 10.1016/j.cpc.2022.108438} {\bibfield  {journal} {\bibinfo  {journal} {Comput. Phys. Commun.}\ }\textbf {\bibinfo {volume} {279}},\ \bibinfo {pages} {108438} (\bibinfo {year} {2022})},\ \Eprint {http://arxiv.org/abs/2206.01457} {arXiv:2206.01457 [physics.comp-ph]} \BibitemShut {NoStop}%
\bibitem [{\citenamefont {Thomas}\ and\ \citenamefont {Ridd}(2004)}]{thomas2004review}%
  \BibitemOpen
  \bibfield  {author} {\bibinfo {author} {\bibfnamefont {S.}~\bibnamefont {Thomas}}\ and\ \bibinfo {author} {\bibfnamefont {P.~V.}\ \bibnamefont {Ridd}},\ }\bibfield  {title} {\enquote {\bibinfo {title} {Review of methods to measure short time scale sediment accumulation},}\ }\href@noop {} {\bibfield  {journal} {\bibinfo  {journal} {Marine Geology}\ }\textbf {\bibinfo {volume} {207}},\ \bibinfo {pages} {95--114} (\bibinfo {year} {2004})}\BibitemShut {NoStop}%
\bibitem [{\citenamefont {Yang}\ \emph {et~al.}(2022)\citenamefont {Yang}, \citenamefont {Luo}, \citenamefont {Yu}, \citenamefont {Zhao}, \citenamefont {Hu}, \citenamefont {Huang}, \citenamefont {Shen}, \citenamefont {Yang}, \citenamefont {Chen},\ and\ \citenamefont {Tang}}]{Yang2022}%
  \BibitemOpen
  \bibfield  {author} {\bibinfo {author} {\bibfnamefont {H.}~\bibnamefont {Yang}}, \bibinfo {author} {\bibfnamefont {G.}~\bibnamefont {Luo}}, \bibinfo {author} {\bibfnamefont {T.}~\bibnamefont {Yu}}, \bibinfo {author} {\bibfnamefont {S.}~\bibnamefont {Zhao}}, \bibinfo {author} {\bibfnamefont {B.}~\bibnamefont {Hu}}, \bibinfo {author} {\bibfnamefont {Z.}~\bibnamefont {Huang}}, \bibinfo {author} {\bibfnamefont {H.}~\bibnamefont {Shen}}, \bibinfo {author} {\bibfnamefont {L.}~\bibnamefont {Yang}}, \bibinfo {author} {\bibfnamefont {Y.}~\bibnamefont {Chen}}, \ and\ \bibinfo {author} {\bibfnamefont {J.}~\bibnamefont {Tang}},\ }\bibfield  {title} {\enquote {\bibinfo {title} {Mugrid: A scintillator detector towards cosmic muon absorption imaging},}\ }\href {\doibase https://doi.org/10.1016/j.nima.2022.167402} {\bibfield  {journal} {\bibinfo  {journal} {Nuclear Instruments and Methods in Physics Research Section A: Accelerators, Spectrometers, Detectors and Associated Equipment}\ }\textbf {\bibinfo {volume} {1042}},\
  \bibinfo {pages} {167402} (\bibinfo {year} {2022})}\BibitemShut {NoStop}%
\bibitem [{\citenamefont {Yao}\ \emph {et~al.}(2006)\citenamefont {Yao} \emph {et~al.}}]{yao2006review}%
  \BibitemOpen
  \bibfield  {author} {\bibinfo {author} {\bibfnamefont {W.-M.}\ \bibnamefont {Yao}} \emph {et~al.},\ }\bibfield  {title} {\enquote {\bibinfo {title} {Review of particle physics},}\ }\href@noop {} {\bibfield  {journal} {\bibinfo  {journal} {Journal of Physics G: Nuclear and Particle Physics}\ }\textbf {\bibinfo {volume} {33}},\ \bibinfo {pages} {1} (\bibinfo {year} {2006})}\BibitemShut {NoStop}%
\end{thebibliography}%

\end{document}